\documentclass[journal]{IEEEtran}
\IEEEoverridecommandlockouts
\usepackage{cite}
\usepackage{amsmath,amssymb,amsfonts}
\usepackage{amsthm,bm}
\usepackage{graphicx}
\usepackage{textcomp}
\usepackage{mathtools}
\def\BibTeX{{\rm B\kern-.05em{\sc i\kern-.025em b}\kern-.08em
    T\kern-.1667em\lower.7ex\hbox{E}\kern-.125emX}}

\usepackage{stackengine}
\usepackage{algorithm}
\usepackage{booktabs}
\usepackage{subfig}
\usepackage{multicol}
\usepackage[noend]{algpseudocode}
\usepackage[table,xcdraw]{xcolor}
\makeatletter
\def\BState{\State\hskip-\ALG@thistlm}
\makeatother

\algnewcommand\algorithmicforeach{\textbf{for each}}
\algdef{S}[FOR]{ForEach}[1]{\algorithmicforeach\ #1\ \algorithmicdo}
    
\begin{document}

\title{\huge Incentive Mechanism Design for Resource Sharing in Collaborative Edge Learning}

\author{}

\author{
    Wei Yang Bryan Lim\thanks{WYB.~Lim and JS. Ng is with Alibaba Group and Alibaba-NTU Joint Research Institute (JRI), Nanyang Technological University (NTU), Singapore. Email: limw0201@e.ntu.edu.sg, s190068@e.ntu.edu.sg. Z.~Xiong is with Alibaba-NTU JRI, and also with School of Computer Science and Engineering (SCSE), NTU, Singapore. Email: zxiong002@e.ntu.edu.sg. D.~Niyato is with SCSE, NTU, Singapore. E-mail: dniyato@ntu.edu.sg. C.~Miao is with SCSE, NTU, Singapore, Alibaba-NTU JRI, and Joint NTU-UBC Research Centre of Excellence in Active Living for the Elderly (LILY), Singapore. Email: ascymiao@ntu.edu.sg. C.~Leung is with LILY and Department of Electrical and Computer Engineering, The University of British Columbia, Canada. Email: cleung@ece.ubc.ca. Q. Yang is with Hong Kong University of Science and Technology, Hong Kong, China. Email: qyang@cse.ust.hk.}, 
    Jer Shyuan Ng,
    Zehui Xiong,
   Dusit Niyato, \IEEEmembership{Fellow, IEEE}, \\
   Cyril Leung, 
    Chunyan Miao,
    Qiang Yang, \IEEEmembership{Fellow, IEEE}

}

\makeatletter
\setlength{\@fptop}{0pt}
\makeatother

\maketitle

\begin{abstract}
In 5G and Beyond networks, Artificial Intelligence applications are expected to be increasingly ubiquitous. This necessitates a paradigm shift from the current cloud-centric model training approach to the Edge Computing based collaborative learning scheme known as edge learning, in which model training is executed at the edge of the network. In this article, we first introduce the principles and technologies of collaborative edge learning. Then, we establish that a successful, scalable implementation of edge learning requires the communication, caching, computation, and learning resources (3C-L) of end devices and edge servers to be leveraged jointly in an efficient manner. However, users may not consent to contribute their resources without receiving adequate compensation. In consideration of the heterogeneity of edge nodes, e.g., in terms of available computation resources, we discuss the challenges of incentive mechanism design to facilitate resource sharing for edge learning. Furthermore, we present a case study involving optimal auction design using Deep Learning to price fresh data contributed for edge learning. The performance evaluation shows the revenue maximizing properties of our proposed auction over the benchmark schemes.
\end{abstract}



\newtheorem{definition}{Definition}
\newtheorem{lemma}{Lemma}
\newtheorem{theorem}{Theorem}

\newtheorem{property}{Property}

\section{Introduction}
\label{sec:intro}

The enhanced perception capabilities of state-of-the-art sensors deployed on, e.g., the Internet of Things (IoT), Unmanned Aerial Vehicles (UAVs), and smart vehicles, have  enabled a wealth of raw data to be captured at the edge of the network today. Empowered with the rise of Artificial Intelligence (AI) based models which outperform conventional hand-engineered methods when there is an abundance of training data, effective models, e.g, for personalized recommendation, facial recognition, and trajectory optimization, have been successfully deployed on end devices, e.g., smart phones. However, with the ubiquity of interconnected end devices, the burden on communication networks necessitates the proposal of Edge Computing as an alternative to current cloud-centric approaches in which raw data has to be transmitted to centralized remote servers for processing. 

The confluence of Edge Computing and AI gives rise to Edge Intelligence, which leverages on the storage, communication, and computation capabilities of end devices and edge servers to enable edge caching, model training, and inference \cite{xu2020survey} closer to where data are produced. Specifically, edge caching refers to the collection and storage of data, edge model training refers to AI model training at the edge, whereas edge inference refers to the deployment of trained models at the edge and end devices for computation of desired outputs, e.g., image classification, given some input data. 


In this article, we refer to the entire pipeline from data collection to model training as \textit{edge learning}. The edge learning offers two main benefits over that of the cloud-centric approach. Firstly, the raw data collected by end devices can be processed locally, without the need for transmission to the remote cloud, thereby reducing communication overhead. Secondly, the computation capabilities of proximal edge servers, e.g., roadside units (RSU), can be leveraged on to complete time-sensitive tasks, e.g., multi-object detection for autonomous driving. 


However, there exist implementation challenges that have to be addressed to enable efficient edge learning. Firstly, privacy regulations that govern the sharing of data, e.g., the General Data Protection Regulation (GDPR), are increasingly stringent \cite{lim2019federated}. Given that edge learning involves the utilization of user\footnote{Note that we use the terms ``users'' and ``end devices'' interchangeably, whereby users are owners of end devices and also data owners.} data for decentralized model training, privacy preserving collaborative learning schemes have to be developed to ensure sustainable participation of users. Secondly, AI models are increasingly complex and over-parameterized \cite{zhang2018shufflenet}. Even though end devices are equipped with improving computation and communication capabilities, they are still much more constrained relative to edge or cloud servers. As such, resource sharing is required at the edge for efficient learning.

In this article, we first introduce the principles and technologies of collaborative edge learning. Then, we argue that successful edge learning requires the sharing of communication, caching, computation, and learning resources ($3$C-L resources) \cite{peltonen20206g} among end devices and edge servers. However, from the perspective of individual users, the consumption of $3$C-L resources is costly, e.g., in terms of device energy cost. Moreover, the $5$G and Beyond networks comprise heterogeneous edge nodes, i.e., users or edge servers, with varying computation and communication capabilities, as well as resource value, e.g., in terms of data quality, all of which are private information. We recognize that incentive mechanism design plays an important role to facilitate sustainable resource sharing for edge learning. Then, we further elaborate the challenges of incentive mechanism design. 

Currently, most incentive mechanism designs for edge learning assume the existence of only one monopolistic model owner. In practice, there may be multiple model owners competing for resources, e.g., user participation in FL. Inspired by the work in \cite{kadota2016minimizing} which proposes the concept of information freshness, i.e., Age of Information (AoI), we present a case study of optimal auction design using Deep Learning to price the fresh data contributed towards model training in FL, when there are multiple competing model owners. The main contributions of this article are as follows:

\begin{itemize}

\item We provide a brief overview of principles and technologies of edge learning.

\item We discuss the motivations behind why $3$C-L resource sharing is required to enable edge learning. Then, we provide a discussion of the challenges of incentive mechanism design in edge learning, as well as the approaches adopted in existing studies.

\item We present a case study in which each user has a certain information freshness and multiple model owners compete for a user's participation in FL. An optimal auction is then designed using Deep Learning.

\end{itemize}

\section{Collaborative Edge Learning}
\label{sec:enable}

In this section, we provide a brief overview of key collaborative edge learning schemes. In general, there are two parties involved in edge learning: i) a model owner, e.g., a developer of AI based applications, and ii) users, e.g., owners of training data collected and stored on-device. There are two main categories of edge learning schemes as follows.

%
%
%
%


\subsection{Federated Learning}
\label{sec:ppc}

Federated Learning (FL) is a decentralized machine learning paradigm proposed in the study in \cite{McMahan2016}. At the beginning of the FL based model training, each user receives a set of identical model parameters from the model owner, i.e., the global model. Each iteration of model training consists of three main steps as illustrated in Fig. \ref{fig:flpro}:
 
\begin{enumerate}
\item \textit{Local model training}: The user trains its received model using locally stored data, e.g., through the Stochastic Gradient Descent (SGD) algorithm to minimize a local loss function.

\item \textit{Parameter transmission}: The user transmits the updated parameters to the model owner.

\item \textit{Global parameter update}: The parameter updates from all users are aggregated and averaged, in the case of the conventional Federated Averaging algorithm  \cite{McMahan2016}, to derive the updated global model.
\end{enumerate}

At the end of each iteration, consisting of the above $3$ steps, the updated global model is transmitted back to the user for the next training round. This process is repeated until a desired model accuracy is achieved. 

In fact, Step $3$ of each iteration is usually conducted in a synchronous manner, i.e., the global parameter update is conducted only after all users have completed the local model training. In contrast, an asynchronous aggregation implies that the global parameter is updated whenever the model owner receives a set of local parameter update from a worker. Most studies of FL presently adopt the synchronous scheme due to the convergence issues of the asynchronous aggregation.

FL avoids the exposure of a user's sensitive raw data to potentially malicious third-party servers. To account for the statistical heterogeneity across users, several studies have emerged with effective frameworks, e.g., through concepts borrowed from multi-task learning in which each user learns personalized models to ensure that high inference accuracy is retained \cite{lim2019federated}.


\subsection{Model Partitioning Based Edge Learning}

FL requires the training and communication of the full set of model parameters between users and the model owner. An alternative approach is the model partitioning based edge learning \cite{shi2020communication} in which the user trains only a partition of the model locally. 

\begin{figure*}
\centering

 \includegraphics[clip, trim=1cm 7cm 1cm 6.5cm,width=0.75\linewidth]{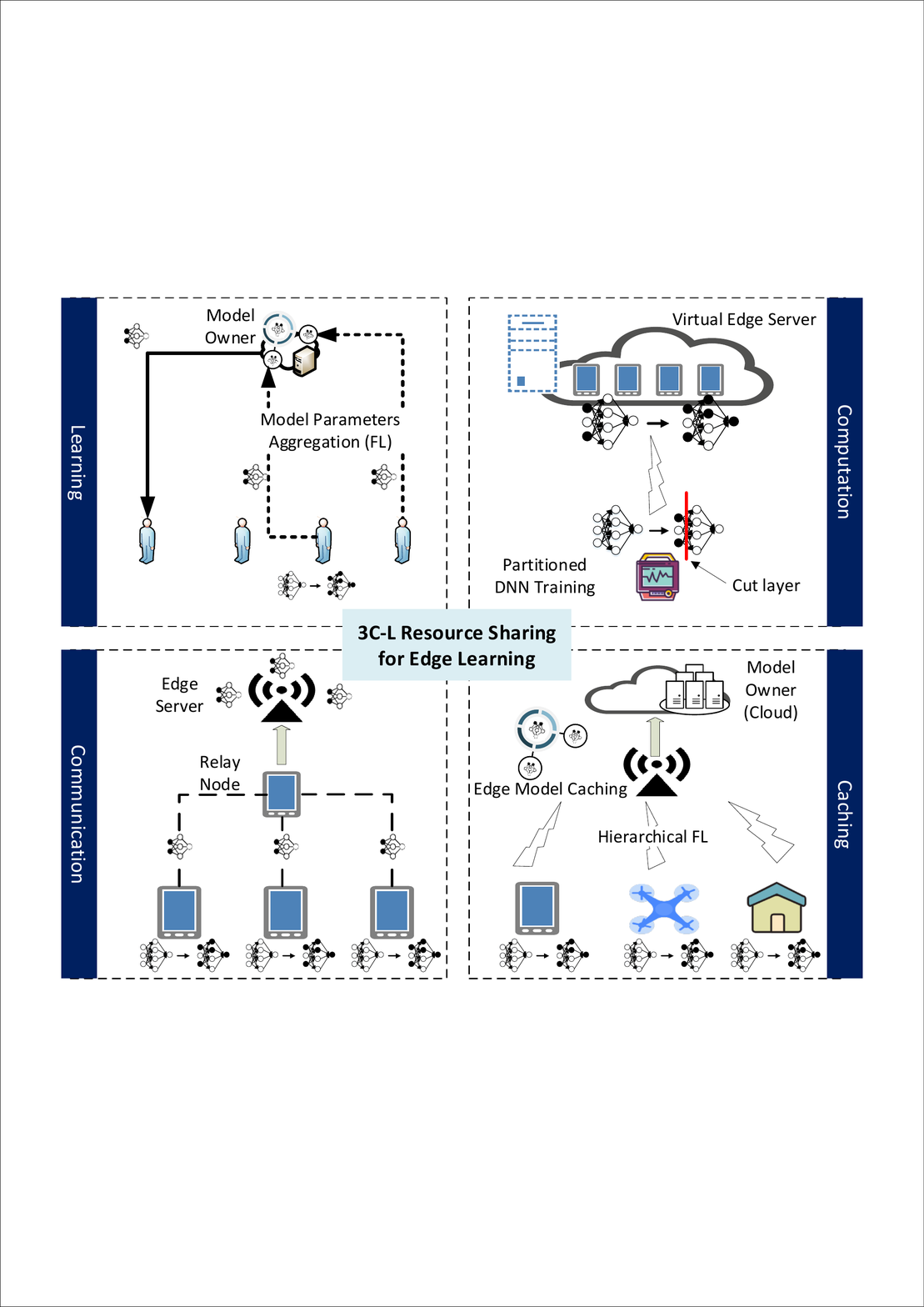}
		\caption{The end-edge-cloud collaboration with $3$C-L resource sharing for edge learning.}
	\label{fig:flpro}

\end{figure*}

Recently, split learning (SL) has been proposed in \cite{vepakomma2018split} whereby each participant only trains the neural network up to a predefined portion, i.e., the cut layer (Fig. \ref{fig:flpro}). Several studies also consider the similar concept of deep neural networks (DNNs) partitioning, in which the shallow layers of the network are trained on-device, whereas the deeper layers are offloaded to the edge servers or cloud for training. In general, the training procedure consists of the following steps:

\begin{enumerate}
\item \textit{Local partitioned model training (forward pass)}: The user trains the DNN up to a predefined layer, e.g., the cut layer.

\item \textit{Computation offloading}: The user transmits the training outputs to an edge server to complete the other layers of training. After a forward pass is completed at the edge or cloud server, the resultant gradients are then backpropagated up to the cut layer and returned to the participants.

\item \textit{Local partitioned model training (backward pass)}: The participants complete the backward pass from the cut layer.
\end{enumerate}

Note that the DNN partitioning can be optimized in consideration of the energy constraints of user devices \cite{kang2017neurosurgeon}. For example, users with computationally-constrained devices can have more layers offloaded to the edge servers for training. On the other hand, users with idle computation resources can complete the training locally, so as to save on communication costs.

Similar to FL, the raw data of users remain locally stored to preserve user privacy, i.e., only the training outputs are transmitted to a third-party server. FL is easier to implement than model partitioning based edge learning. In FL, the model owner merely has to perform simple aggregation, e.g., through Federated Averaging, of the collected local parameters. However, the model partitioning based edge learning is suitable in applications in which a large DNN model is involved and it may be infeasible for computationally-constrained users to train completely on-device.

%
%



\section{Resource Sharing at the Edge For Efficient Learning}
\label{sec:resourcesharing}

The key to collaborative edge learning is resource sharing, i.e., the contribution of communication, computation, caching, and learning resources ($3$C-L) from end devices and edge servers to enable model training at the edge of the network. It is envisioned that in $5$G networks and beyond, the $3$C-L resources of end devices and edge users can be virtualized as a common resource pool to satisfy the user requirements, e.g., Quality of Experience (QoE), and support collaborative edge learning for the development of ubiquitous AI applications \cite{peltonen20206g}. In this section, we discuss the $3$C-L resource sharing at the edge for collborative learning.

\subsection{Communication}

The collaborative schemes discussed in Section \ref{sec:enable} require communication of training outputs to the aggregating server. However, users from remote regions with limited bandwidth may have high dropout rates which disrupt the model training process. To reduce dropout rates and enable the inclusion of a wider group of contributing users for the model training, an intermediate aggregation at proximal edge servers can be adopted. In hierarchical FL \cite{lim2019federated}, local model parameters can first be aggregated on edge servers (Fig. \ref{fig:flpro}). After several rounds of intermediate aggregation, the parameters are then transmitted from the edge servers to the cloud for global aggregation. This can reduce the costs of communication with the remote cloud. 

However, certain end devices may still lack the required connectivity for update transmission to edge servers. In such cases, cooperative communication can be adopted in which end devices can first be grouped into clusters. Then, the trained parameters from each cluster are transmitted to a relay node situated between the cluster and the edge server so as to increase the network connectivity and capacity \cite{xu2020survey}.

\subsection{Computation}

In model partitioning based edge learning, the deeper layers of the DNN are offloaded, e.g., to cloud servers, for model training given the hardware constraints of end devices.

To leverage on the available computation resources at the edge of the network, device-to-edge (D2E) offloading can be implemented with a lower communication cost incurred. Moreover, clusters of end devices, e.g., stationary vehicles \cite{zhou2019computation}, can potentially serve as virtual edge servers for device-to-device (D2D) computation offloading \cite{xu2020survey}.

\subsection{Caching}

Traditionally, the caching of popular files is implemented, e.g., in multimedia streaming, to manage user request and reduce communication and computation redundancy. With the ubiquity of AI in our daily lives, AI models can now be cached in edge servers \cite{xu2020survey,lim2019federated} for communication efficient edge learning. For example, the aforementioned hierarchical FL involves the caching of intermediate models at the edge (Fig. \ref{fig:flpro}). At the beginning of each training iteration, the cached model is transmitted to each end user for local model training. Then, the cached model is updated through intermediate aggregation, and communicated to the users again for another iteration of training. The proximity of the edge server reduces training latency and user dropout rates. 
%

\subsection{Learning}

The proximity of edge servers enable efficient crowdsensing whereby training data can be aggregated and validated at the edge with reduced latency \cite{zhou2018robust}. Beyond crowdsensing, a new paradigm called \textit{crowd learning} can be enabled with collaborative edge learning. Amid data privacy concerns, the privacy preserving properties of, e.g., FL, will encourage greater participation in model training. Given that representation learning models perform well when large quantities of training data and abundant computation resource are available \cite{lim2019federated}, crowd learning can support the training of models that generalize well by leveraging on the sensing, computation, and communication resources of end users to train a shared model. In fact, the next-word-prediction model of Gboard\footnote{gboard.app.goo.gl/get} has been successfully trained using FL, whereas SL has been considered for healthcare applications \cite{vepakomma2018split}.

\section{Incentive Mechanism Design}

In Section \ref{sec:resourcesharing}, we discuss how the $3$C-L resources of end devices and edge servers can be leveraged to facilitate efficient edge learning. However, users may not consent to share their resources without receiving compensation or reward for the corresponding costs incurred, e.g., energy cost due to model training. Thus, there is a need for incentive mechanism \cite{peltonen20206g} to facilitate resource sharing for edge learning. For convenience, we refer to the model owner as the buyer (of $3$C-L resources) and the user (who is the owner of the resource) as the seller. We also tag each seller with a \textit{type} according to the underlying value of the seller's resource, e.g., high type sellers with more training data available are more desirable than low type sellers with less training data available. 

In the following, we discuss key challenges of incentive mechanism design:

\begin{itemize}

\item \textit{Incentive mismatch:} There is a fundamental difference in the objectives between buyers and sellers in the resource sharing market for edge learning. On one hand, buyers wish to minimize costs in providing incentives while maximizing the quality of the resources acquired from sellers. On the other hand, sellers wish to receive the maximum compensation for their resources. Incentive designs are required to achieve a balance between these conflicting goals.

\item \textit{Network heterogeneity and information asymmetry:} The $5$G and Beyond networks are envisioned to consist of thousands of heterogeneous end devices and edge servers, with different capabilities and resource values.  However, as a result of the privacy-preserving properties of collaborative learning schemes, it is not possible for the resource buyer to evaluate a resource, e.g., computation capability, directly. A well-designed incentive mechanism should elicit the truthful revelation of seller types.

\item \textit{Dynamic Conditions:} Conventional incentive mechanism designs are based on analytical methods. However, given the growing complexity and capacity of future networks, the dynamic conditions in the network cannot be completely modeled or solved with standard approaches \cite{wang2019edge}. As a result, DNNs are being increasingly adopted for the design of learning-based incentive mechanisms \cite{zhan2020learning}.

\end{itemize}

Some most commonly used incentive tools considered in existing studies are as follows:

\begin{itemize}

\item \textit{Contract Theory:} In contract theory, contract bundles consisting of resource contribution-reward pairs are designed for different seller types. The contracts are designed such that the incentive compatibility constraint is met, i.e., each utility maximizing seller only chooses the contract designed for its own type. This ensures that rational sellers do not misreport their types. 

\item \textit{Game Theory:} The Stackelberg game is particularly useful in modeling edge learning schemes, with the model owner (buyer) assuming the role of a leader whereas the users (sellers) are followers. Specifically, the game consists of two stages. In the first stage, the leader announces a reward. In the second stage, the followers determine the level of resource contribution, e.g., computation resource contributed for edge offloading, in response to the expected reward in the first stage, as well as the actions of other followers \cite{lim2019federated}. 

\item \textit{Auction Theory:} In an optimal auction design, a buyer only submits a truthful bid for a seller's resource, i.e., there is no incentive to misreport its valuation of a resource. For reverse auctions, sellers bid for the prices at which they are willing to sell their resource at, e.g., participation in FL based model training. In contrast to procedures in the contract and game theory approaches, the reverse auction approach \cite{jiao2019toward} enables users to actively report their types.

\end{itemize}

Most incentive mechanism design studies in edge learning assume the existence of only one monopolistic buyer. In practice, there may be multiple buyers, e.g., rival companies, each seeking to develop an AI application for their own purposes. For this scenario, we present a case study in which an auction is held among \textit{multiple} model owners who compete for a user's resource.


%
%

\section{Optimal Auction with Deep Learning}

\begin{figure}
\includegraphics[clip, trim=0cm 0cm 0cm 0cm, width=\columnwidth]{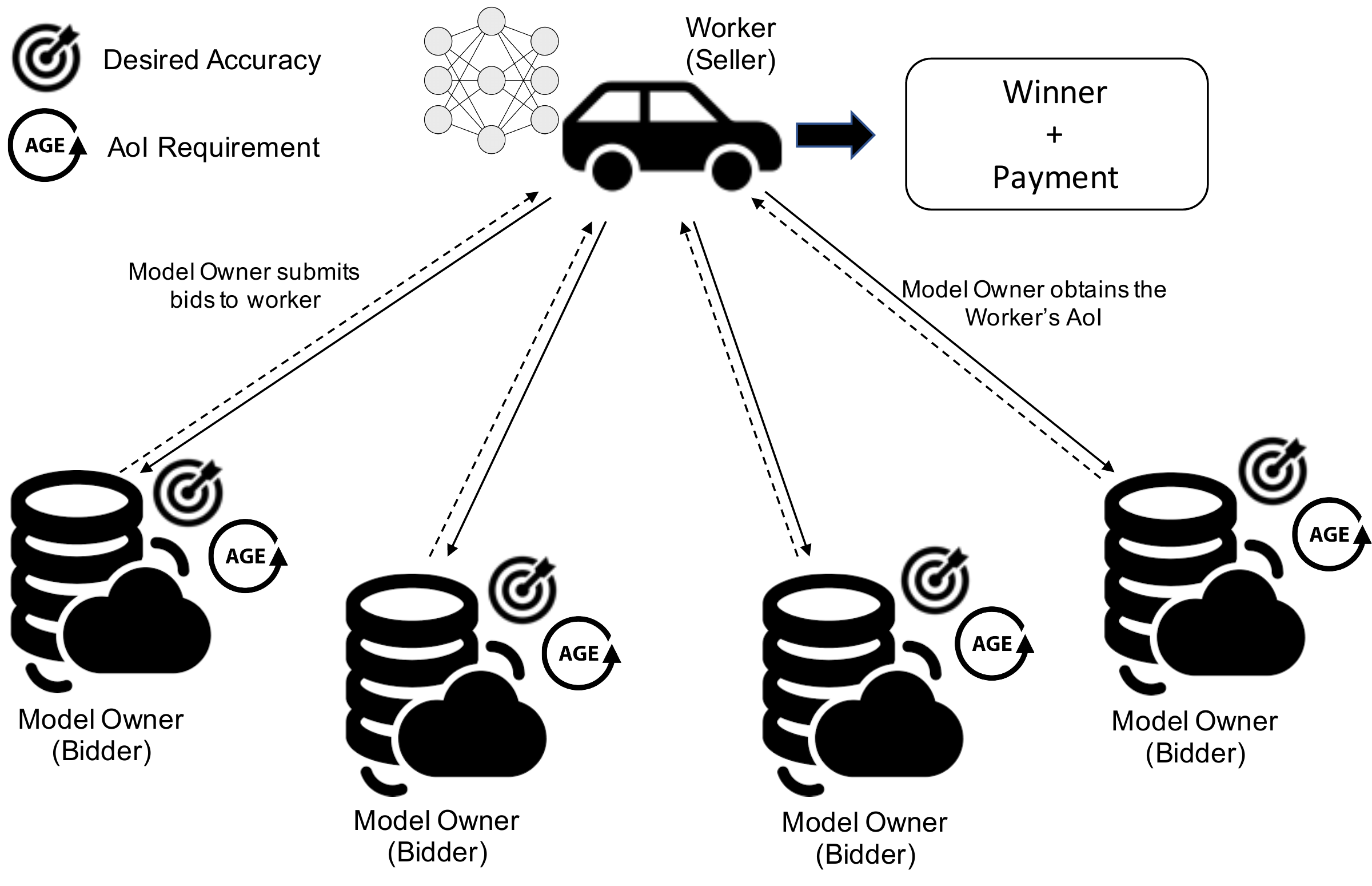}
\caption{System Diagram of the Optimal Auction in FL.}
\label{system}
\end{figure}

In this section, we present a case study in which an auction is held among \textit{multiple} model owners (bidders) to buy the resources of a worker (sellers in Fig. \ref{system}) in the FL based model training. We then use a Deep Learning approach to ensure truthful reporting of the model owner's valuation amid information asymmetry. Note that the proposed mechanism can be straightforwardly applied to other model partitioning based edge learning methods such as split learning.

At any point of time, each worker can only participate in the FL training initiated by one model owner, i.e., there can only be a single winner of the auction. Workers that are found to sell their services to more than one model owner simultaneously are penalized, e.g., by being barred from participating in future training instances. Moreover, to preserve the privacy of workers, the model training is implemented locally as discussed in Section \ref{sec:ppc}.

\begin{figure*}
\centering
\begin{multicols}{2}

\includegraphics[clip, trim=1cm 0cm 0cm 2.5cm, width=\columnwidth]{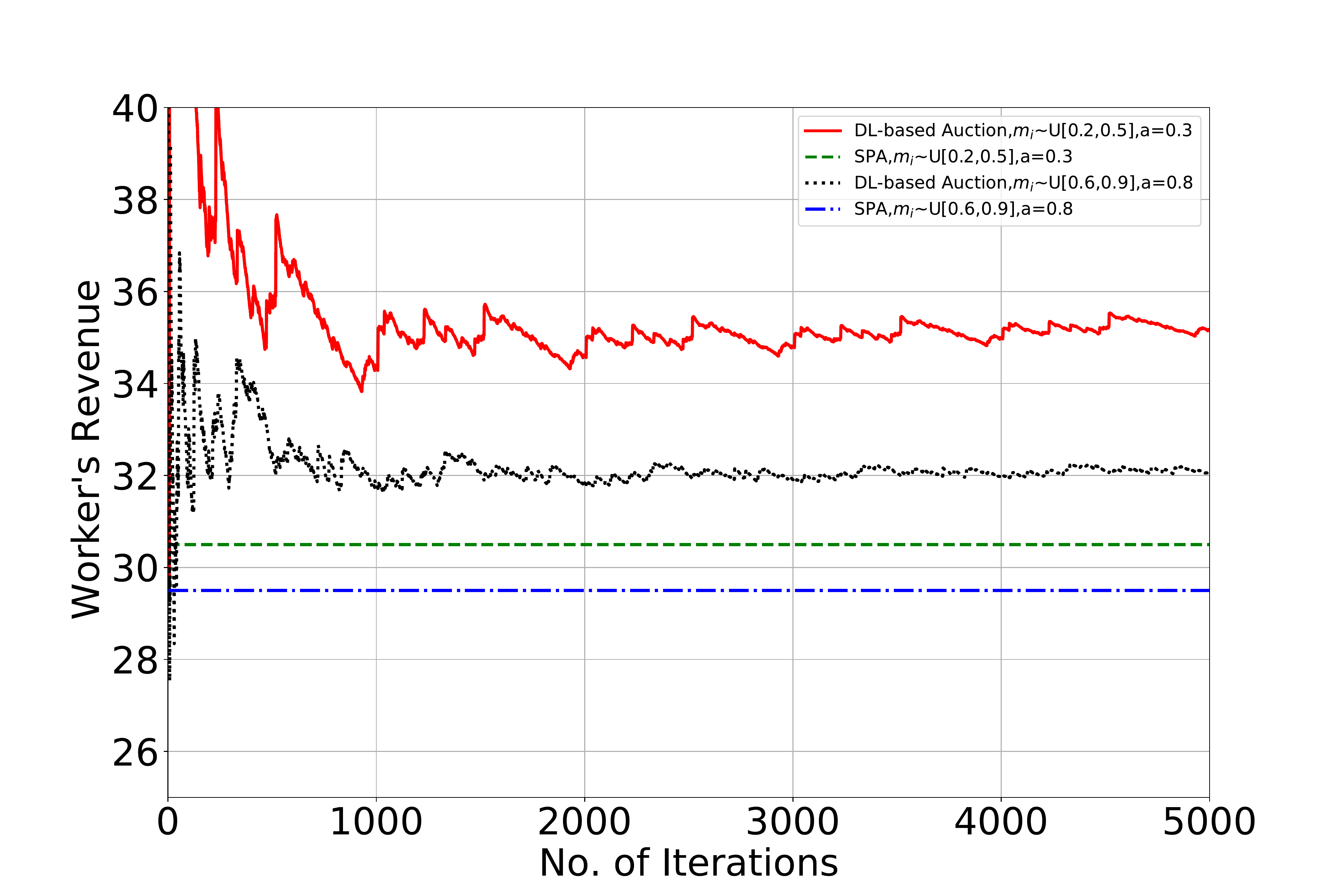}
\caption{Comparison of Deep Learning-based auction and SPA.}
\label{comparespa}

\includegraphics[clip, trim=1cm 0cm 0cm 2.5cm, width=\columnwidth]{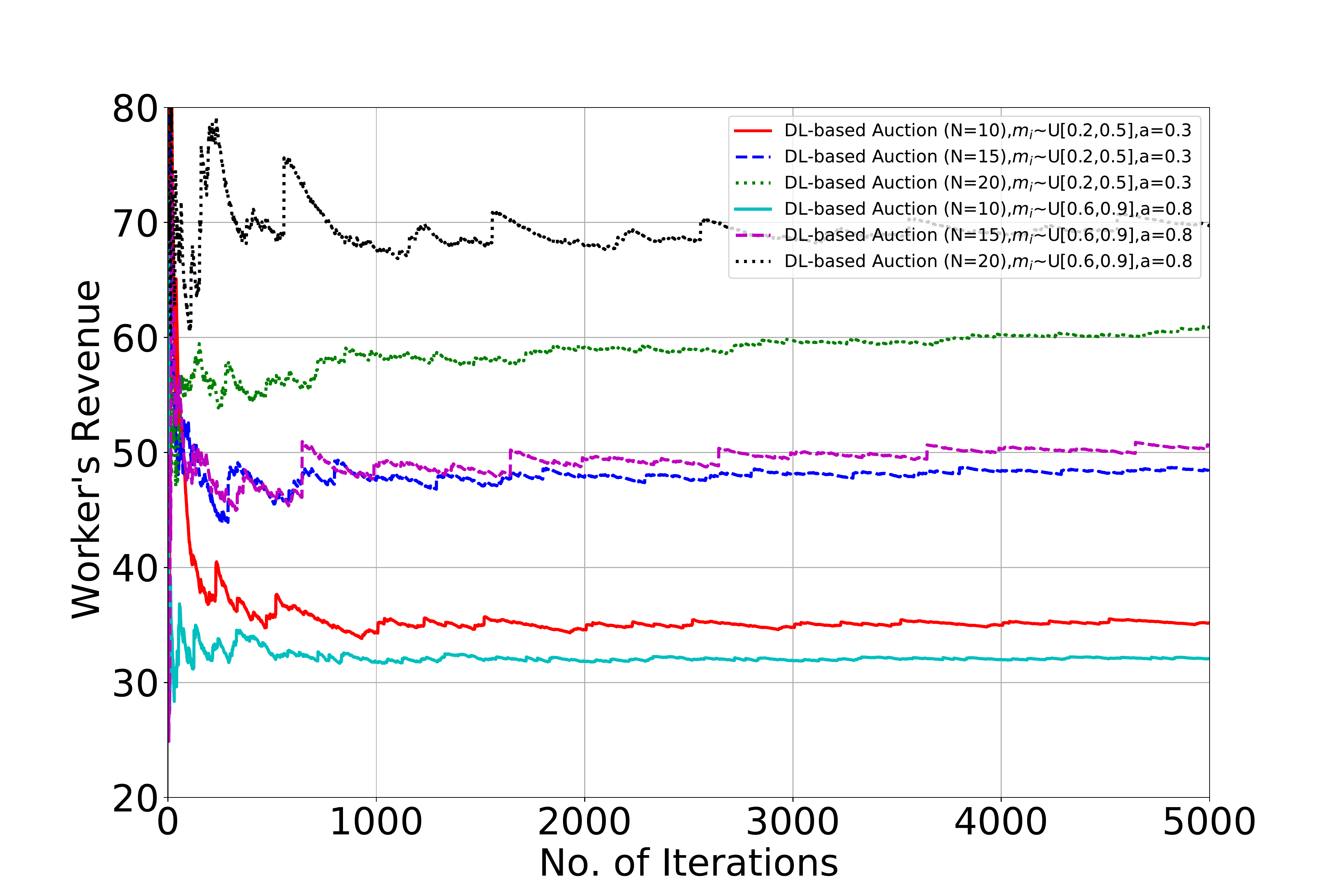}
\caption{Simulating revenue under varying AoI requirements.}
\label{matchingrequirements}

    \end{multicols}
    
\end{figure*}

\subsection{Auction Framework}

We consider a system model in which a worker's data are associated with a particular AoI \cite{kadota2016minimizing}, i.e., the time elapsed between the collection of data by the worker to the completion of the FL model training. For example, a worker which updates its data more frequently has a lower AoI, i.e., the data it uses for model training is relatively fresh. 
 
If the model owner prefers data with low AoI values, e.g., for the development of navigation systems in autonomous vehicles, it has an incentive to pay a higher price to the worker with low AoI. In contrast, if the model owner does not value fresh data, e.g., for the development of a location-based recommender system in which user preferences are only slowly time-varying, it will not pay a high premium for data with low AoI. Specifically, the bid value of the model owner is inversely proportional to the AoI associated with a worker. 


To maximize the revenue of the worker and ensure that each worker is allocated to the model owner that values its corresponding AoI the most, the worker allocation problem can be modeled as a single-item auction. Through the auction process, a winning model owner and the corresponding payment for the services of the worker are determined. An optimal auction  has two characteristics:
\begin{enumerate}
\item \emph{Individual Rationality (IR): }By participating in the auction, the model owners receive non-negative payoff.
\item \emph{Incentive Compatibility (IC): }There is no incentive for the model owners to submit bids other than their true valuations, i.e., the bidders always bid truthfully.
\end{enumerate}
The traditional first-price auction, in which the highest bidder pays the exact bid it submits, maximizes the revenue of the seller but does not ensure that the bidders submit their true valuation. The second-price auction (SPA), in which the highest bidder pays the price offered by the second highest bidder, ensures that the bidders submit their true valuations but does not maximize the revenue of the seller. In order to ensure that both conditions of truthfulness and revenue maximization of the seller are satisfied, an optimal auction is designed using the Deep Learning approach \cite{luong2018optimal}.

\begin{figure*}
\centering
\begin{multicols}{2}

\includegraphics[clip, trim=1cm 0cm 0cm 2.5cm, width=\columnwidth]{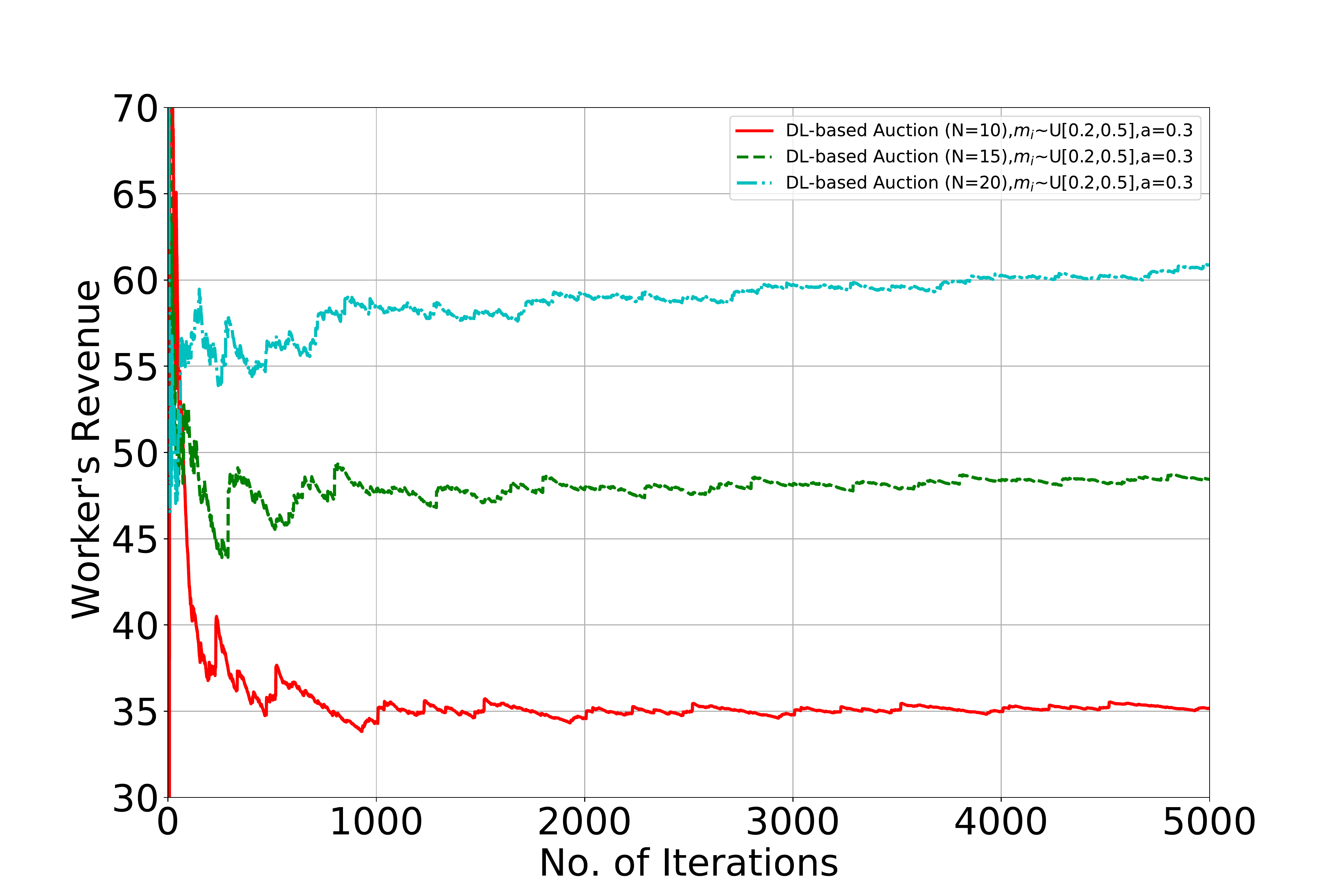}
\caption{Simulating revenue given different number of participating bidders.}
\label{diffnumbidders}

\includegraphics[clip, trim=1cm 0cm 0cm 2.5cm, width=\columnwidth]{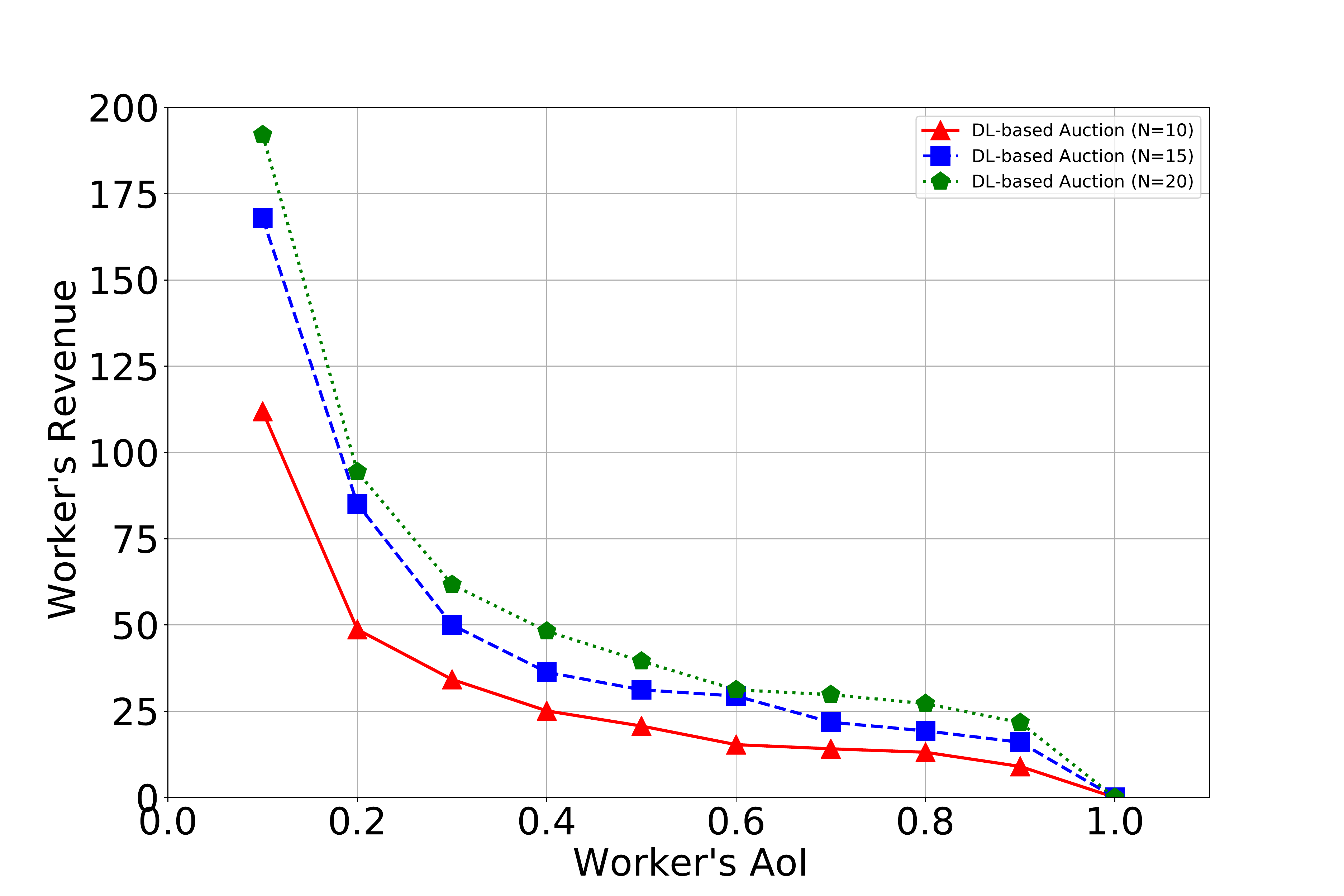}
\caption{Revenue vs. worker AoI.}
\label{revaoi}

    \end{multicols}
    
\end{figure*}

\subsection{Deep Learning-Based Optimal Auction}

In the following, we describe the Deep Learning-based auction algorithm \cite{luong2018optimal}. The bids of the model owner are first transformed to derive the transformed bids, which meet the IR and IC requirements. Then, a winner and the corresponding payment are determined. 

\begin{enumerate}

\item \emph{Monotonic Transform Functions: }To ensure that the winner determination and the corresponding payment satisfy the IR and IC requirements, the transform functions have to be strictly monotonically increasing. The  precise form of the transform function is derived using a two-layer feed forward network trained to minimize the loss of the network, i.e., the negated revenue of the worker. In other words, this procedure is equivalent to the maximization of the revenue of the worker. 
\item \emph{Winner Determination: }The winner is determined based on the SPA with zero reserve price allocation rule. This allocation rule maps the transformed bids to a vector of assignment probabilities, i.e., winning probabilities, using an approximated softmax function and a dummy input. If the transformed bid is positive, the model owner with the highest transformed bid wins the auction. Otherwise, the worker does not participate in the FL model training with any model owner.
\item \emph{Conditional Payment Price: }To derive the required payment to be made by the winner, the payment prices of all bidders are first calculated, followed by the conditional price of the winning model owner. The payment prices of the model owners are determined using a \emph{ReLU} activation function whereas the conditional price of the winning model owner is determined  using the monotonic inverse transform function.
\end{enumerate}

\section{Performance Evaluation}
In this section, we evaluate the performance of the Deep Learning-based auction. For comparison, the classic SPA is chosen as the baseline scheme. The TensorFlow Deep Learning library is used to implement the optimal auction design. In our simulations, the AoI of the worker and the AoI preference of the model owners follow uniform distributions, e.g., a preference for lower AoI value implies that the buyer values fresh information more.

%
%
%
%
%
%
%
%

%

Figure \ref{comparespa} shows that the Deep Learning approach allows the worker to earn higher revenue than that of the traditional SPA scheme. The reason is that the SPA scheme guarantees IC, but does not guarantee that the revenue of the seller is  maximized. 



To illustrate the impact of model owners' preference for AoI on the worker's revenue, we consider the scenario in which the worker has varying AoI, i.e., $0.3$ and $0.8$. From Fig. \ref{matchingrequirements}, the worker's revenue is greatly increased when the worker's AoI is low and falls within the range of AoI requirements by the model owners. This serves as a compensation for the greater data processing cost incurred by the worker to keep the AoI low. Note that if the worker has an AoI lower than the model owner requirement, there will be no winner in the auction. Intuitively, model owners with AoI requirements above the worker AoI has no incentive to pay a higher price for fresh data.

Furthermore, we consider the worker's revenue under different number, $N$, of bidders in Fig. \ref{diffnumbidders} and \ref{revaoi}, i.e., $N=10, N=15$ and $N=20$. As $N$ increases, the worker's revenue increases. The competition among the model owners increases with $N$. As such, model owners have greater incentive to submit higher bids in order to win the auction. As a result, the revenue of the worker increases. Moreover, workers with fresher data receive higher revenue in compensation of the greater data collection cost incurred.

\section{Conclusion and Future Research Directions}

In this article, we first reviewed the principles and technologies in collaborative edge learning, followed by an overview of incentive mechanism design to facilitate the $3$C-L resource sharing at the edge for efficient learning. We then presented a case study for a scenario in which multiple buyers compete for fresh data from a single seller. The performance evaluation of a Deep Learning based auction shows the seller revenue maximization properties of our designed auction. 

For our future research directions, we aim to consider:
\begin{itemize}

\item \textit{Multiple bidder-multiple seller auction for $3$C-L resources}: Currently, we have assumed the multiple bidder-single seller auction framework. With multiple sellers involved, the resulting competition among sellers may drive the seller revenue downwards.

\item  \textit{Dynamic resource availability of sellers and buyers}: In some cases, edge learning is conducted over a period of time, during which the sellers may have varying resources, e.g., data quality, across time. In our proposed mechanism, as well as most existing works, the dynamic resource value is not considered.

\item \textit{Collusion and coalition formation among model owners}: In practice, model owners may collude to bid for a resource. This may in turn, affect the bidding strategies and seller's revenue.

\end{itemize}

\bibliographystyle{IEEEtran}
\bibliography{fl-uav}

\end{document}